\documentclass[prd,showpacs]{revtex4}
\usepackage{epsfig}
\usepackage{amsmath}
\usepackage{amssymb}

\begin{document}

\title{ Study of $B_{c}\to J/\psi K$  decays
in the perturbative QCD approach}

\author{Xian-Qiao Yu\footnote{yuxq@swu.edu.cn}, Xiu-Li Zhou}

\affiliation{
 {\it \small School of Physical Science and Technology, Southwest
 University, Chongqing 400715, China}}

\begin{abstract}
  In this note, we calculate the
branching ratio of $B_c\rightarrow J/\psi K$ in the framework of
perturbative QCD approach based on $k_T$ factorization. This decay
can occur only via tree level diagrams in the Standard Model. We
find that the branching ratio of $B_c\rightarrow J/\psi K $ is
about $(1-3)\times10^{-4}$. The large branching ratio and the
clear signals of the final states make the measurement of
$B_c\rightarrow J/\psi K $ easily at LHC-b experiments.
\end{abstract}

 \maketitle

The $B$ meson rare decays provide a good place for testing the
Standard Model (SM), studying $CP$ violation and looking for
possible new physics beyond the SM. The theoretical studies of
$B_{u,d}$ mesons decays have been studied widely in the
literature, which are strongly supported by the detectors at the
$e^{+}e^{-}$ colliders, such as the CLEO, BaBar and Belle. At the
era of the Large Hadron Collider(LHC), there is still a room for B
physics. LHC beauty experiments(LHCb) will extend the B-physics
results from the B factories by investigating decays of heavier B
hadrons, such as $B_s$ and $B_c$ mesons. It is estimated that
about $5\times10^{10}$ $B_c$ mesons can be produced per year at
LHC\cite{LHC,AT}, so the studies of $B_{c}$ meson rare decays are
necessary in the next a few years, it will highlight the
advantages of B physics.

In this paper, we study the rare decays $B_{c}\to J/\psi K$ in the
Perturbative QCD approach (PQCD) \cite{LY}. In SM, $B_{c}\to
J/\psi K$ decays occur through only the tree level diagrams and so
there is no CP violation in this channel. Therefore, the
measurements of this channel might provide a ground for
investigating new physics effects.

For the decay $B_{c}\to J/\psi K$, the related effective
Hamiltonian is given by \cite{BBL}
\begin{equation}
 H_\mathrm{eff} = \frac{G_F}{\sqrt{2}}V_{us}V_{cb}^*\left\{
C_1(\mu) O_1(\mu) + C_2(\mu) O_2(\mu)\right\}+H.c.,\label{hami}
\end{equation}
 where $C_{i}(\mu)$ are Wilson coefficients at the
  renormalization scale $\mu$ and $O_{i}$ are the local four-quark operators
\begin{equation}\begin{array}{ll}
  O_1 = (\bar{b}_ic_j)_{V-A}(\bar{u}_js_i)_{V-A},  &
  O_2 = (\bar{b}_ic_i)_{V-A} (\bar{u}_js_j)_{V-A}. \label{eq:effectiv}
 \end{array}
\end{equation}
Here $i$ and $j$ are $SU(3)$ color indices. Then the calculation
of decay amplitude is to evaluate the hadronic matrix elements of
the local operators.

 In the PQCD approach, the decay amplitude can be written as:
\begin{equation}
 \mbox{Amplitude}
\sim \int\!\! d^4k_1 d^4k_2 d^4k_3\ \mathrm{Tr} \bigl[ C(t)
\Phi_{B_c}(k_1) \Phi_{\psi}(k_2) \Phi_K(k_3) H(k_1,k_2,k_3, t)
\bigr]e^{-S(t)}. \label{eq:convolution1}
\end{equation}
In our following calculations, the Wilson coefficient $C(t)$,
Sudakov factor $S_{i}(t)(i=B_c, J/\psi, K)$ and the
non-perturbative but universal wave function $\Phi_{i}$ can be
found in the Refs. \cite{LUY,ZY,SDY,CDL,YLL,ATM}. The hard part
$H$ are channel dependent but fortunately perturbative calculable,
which will be shown below.

\begin{figure}[htb]
\vspace{0.5cm}
\begin{center}
\includegraphics[scale=0.75]{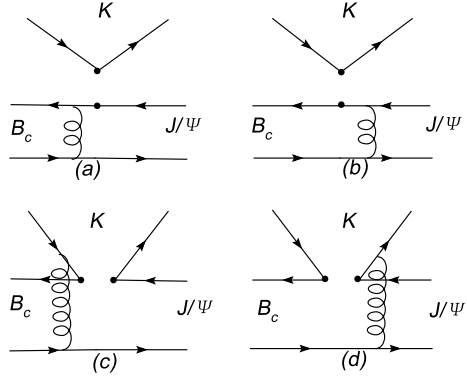}
\caption{The lowest order diagrams for $B_{c} \to J/\psi K$
decay.} \label{figure:Fig1}
\end{center}
\end{figure}

 Fig.~\ref{figure:Fig1} shows the lowest order
Feynman diagrams to be calculated in the PQCD approach where (a)
and (b) are factorizable topology, (c) and (d) are nonfactorizable
topology. After a straightforward calculation using the PQCD
formalism Eq.(\ref{eq:convolution1}), we obtain the sum
contributions of (a) and (b)

\begin{eqnarray}
M_a[C] &=&\frac{8f_{B_{c}}}{\sqrt{2N_{c}}}\pi C_{F}
M_{B_{c}}^{2}\int_{0}^{1}dx_{2}\int_{0}^{\infty}b_1db_1b_{2}db_{2}
\times\{[r_br(1-r^2)\phi_{\psi}^{t}(x_{2},b_2)
-2r(1-r^{2})x_2\phi_{\psi}^{t}(x_{2},b_2)\nonumber\\
&&+(1-r^2)x_2\phi_{\psi}^{L}(x_{2},b_2)
-2r_{b}(1-r^2)\phi_{\psi}^{L}(x_{2},b_2)]
\alpha_{s}(t_{a}^{1})h_{a}^{(1)}(x_{2},b_1,b_{2})\exp[-S_{B_c}(t_{a}^{1})-S_{J/\psi}(t_{a}^{1})]
C(t_{a}^{1})\nonumber\\
&&+[r^2(1-r^2)(r_c-1)\phi_{\psi}^{L}(x_{2},b_2)-r_c(1-r^2)\phi_{\psi}^{L}(x_{2},b_2)]
\alpha_{s}(t_{a}^{2})h_{a}^{(2)}(x_{2},b_1,b_{2})\nonumber\\
&&\times\exp[-S_{B_c}(t_{a}^{2})-S_{J/\psi}(t_{a}^{2})]
C(t_{a}^{2})\},\label{ma}
\end{eqnarray}
where $r=M_{J/\psi}/M_{B_c}, r_{b}=M_b/M_{B_c}, r_c=M_c/M_{B_c}$
and
 \begin{eqnarray}
\nonumber t_a^{1} &=& \mathrm{max}(M_B
\sqrt{(1-x_{2})(r_c-r^2)},M_B \sqrt{|r_{b}^{2}-(1-r^2)x_2|},
 1/b_1,1/b_2), \\
 t_a^{2} &=& \mathrm{max}(M_B \sqrt{r_{c}^{2}+r_c-r^2},
 1/b_1,1/b_2).\label{maxt}
 \end{eqnarray}
  The functions
\begin{align}
& h_{a}^{(1)}(x_2,b_1,b_2) = S_{t}(x_2)K_{0}(M_{B}\sqrt{(1-x_2)(r_c-r^2)}b_{1})\nonumber \\
& \times
 [\theta(b_2-b_1)\theta(r_{b}^{2}-(1-r^{2})x_{2})I_0(M_B\sqrt{r_{b}^2-(1-r^2)x_2}b_1)\nonumber \\
 &\times
 \mathrm{K}_0(M_B\sqrt{r_{b}^2-(1-r^2)x_2}b_2)
 +(b_1\leftrightarrow b_2)],
\end{align}
\begin{align}
& h_{a}^{(2)}(x_2,b_1,b_2) = S_{t}(r_c)K_{0}(M_{B}\sqrt{(1-x_2)(r_c-r^2)}b_{2})\nonumber \\
& \times
\bigl[\theta(b_2-b_1)I_{0}(M_{B}\sqrt{r_c^{2}+r_c-r^2}b_1)K_{0}(M_{B}\sqrt{r_{c}^{2}+r_c-r^2}b_2)+(b_1
\leftrightarrow b_2)\bigr],
 \label{eq:propagator1}
\end{align}
come from the Fourier transformation of propagators of virtual
quark and gluon in the hard part calculations. For the
non-factorizable diagrams (c) and (d), all three meson wave
functions are involved. Their total contribution is:
\begin{eqnarray}
M_c[C]  &=& -\frac{16}{N_{c}}\pi C_{F}
f_{B_{c}}M_{B_{c}}^{2}\int_{0}^{1}dx_{2}dx_3\int_{0}^{\infty}b_{2}db_{2}b_3db_3\phi_{K}^{A}(x_3) \nonumber\\
 &&\times\{[(1-r^2)(1-r^2-r_c-x_3+r^{2}r_c+r^{2}x_3)\phi_{\psi}^{L}(x_{2},b_2)+r(1-r^2)(1-r_c-x_2)\phi_{\psi}^{t}(x_{2},b_2)]\nonumber\\
&&
\alpha_{s}(t_{c}^{1})h_{c}^{(1)}(x_2,x_3,b_{2},b_3)\exp[-S_{B_c}(t_{c}^{1})-S_{J/\psi}(t_{c}^{1})-S_{K}(t_{c}^{1})]
C(t_{c}^{1})+[(r^2-1)(1+r^2-2r_c-x_2+x_3 \nonumber\\
&&-r^2x_2-r^2x_3)\phi_{\psi}^{L}(x_{2},b_2)
+r(1-r^2)(1-r_c-x_2)\phi_{\psi}^{t}(x_{2},b_2)]
\alpha_{s}(t_{c}^{2})h_{c}^{(2)}(x_2,x_3,b_{2},b_3)\nonumber\\
&&\exp[-S_{B_c}(t_{c}^{2})-S_{J/\psi}(t_{c}^{2})-S_{K}(t_{c}^{2})]
C(t_{c}^{2})\},\label{mc}
\end{eqnarray}
where
\begin{eqnarray}
\nonumber t_c^{1} &=& \mathrm{max}(M_B
\sqrt{|F^2_{(1)}|},M_B\sqrt{(r_c-r^2)(1-x_2)},
 1/b_2,1/b_3), \\
 t_c^{2} &=& \mathrm{max}(M_B
\sqrt{|F^2_{(2)}|},M_B\sqrt{(r_c-r^2)(1-x_2)},
 1/b_2,1/b_3),
 \end{eqnarray}
\begin{align}
& h_{c}^{(j)}(x_2,x_3,b_2,b_3) = \nonumber \\
&
\biggl\{\theta(b_2-b_3)I_{0}(M_B\sqrt{(1-x_2)(r_c-r^2)}b_3)K_{0}(M_B\sqrt{(1-x_2)(r_c-r^2)}b_2)
\nonumber \\
& \qquad\qquad\qquad\qquad + (b_2\leftrightarrow b_3) \biggr\}
 \times\left(
\begin{matrix}
 \mathrm{K}_0(M_B F_{(j)} b_3), & \text{for}\quad F^2_{(j)}>0 \\
 \frac{\pi i}{2} \mathrm{H}_0^{(1)}(M_B\sqrt{|F^2_{(j)}|}\ b_3), &
 \text{for}\quad F^2_{(j)}<0
\end{matrix}\right),
\label{eq:propagator2}
\end{align}
 and $F_{(j)}$'s are defined by
\begin{eqnarray}
 F^2_{(1)} = (1-r_c-x_3+r^2x_3)(x_2-1),\nonumber \\
F^2_{(2)} = (1-x_2)(r_c-r^2-x_3+r^2x_3).
\end{eqnarray}

The total decay amplitude is then
\begin{align}
{ \cal A} (B_{c}^{+} \to J/\psi K^{+})
=V_{us}V_{cb}^{*}\left[C_{1}M_c+(\frac{1}{3}C_{1}+C_{2})M_{a}f_{K}\right],
\end{align}
and the decay width is expressed as

\begin{equation}
 \Gamma(B_{c}^{+} \to J/\psi K^{+}) = \frac{G_F^2
 M_{B_c}^{3}}{128\pi}(1-r^2)
\left|{\cal A} (B_{c}^{+} \to J/\psi K^{+})\right|^2.
\label{eq:width3}
\end{equation}

The following parameters have been used in our numerical
calculation  \cite{PDG,SI,TWQCD,Melic,Hashimoto}:
\begin{gather}
\nonumber  M_{B_c} = 6.286 \pm 0.005\mbox{GeV}, M_{b} = 5.2
\mbox{GeV}, M_{J/\psi} = 3.097 \mbox{GeV}, M_{c} = 1.82
\mbox{GeV},
 \\ \nonumber  f_{B_c} = 489 \pm4\mbox{MeV},
 f_{K} = 160 \mbox{MeV}, f_{J/\psi} = 405\pm14 \mbox{MeV},
 \\  \tau_{B_c}=(0.46\pm0.07)\times 10^{-12}\mbox{s},
  |V_{us}|=0.2257\pm0.0021, |V_{cb}|=(41.6\pm0.6)\times 10^{-3}.
\label{eq:shapewv}
\end{gather}
If not specified, we shall take their central values as the
default input. We have taken the constituent quark masses $M_{b} =
5.2GeV$ and $M_{c} = 1.82GeV$ from the ISGW2 model\cite{SI}. As
noted in Ref.\cite{Anisimov}, this choice for $M_b$ and $M_c$
satisfies approximately the relation $M_b=M_c+3.4GeV$, which is
consistent with the well known formula relating the pole masses
$M_{b,pole}$ and $M_{c,pole}$ in the Heavy Quark Effective Theory.
We also update some parameters used in previous works by taking
the values from the latest Particle Data Group publication and
lattice QCD simulations.

Our numerical analysis shows that
$|C_1M_c/((\frac{1}{3}C_1+C_2)M_af_k)|=17\%$, which means that the
dominated contributions to the branching ratio of $B_c\rightarrow
J/\psi K$ decays come from the factorizable topology[(a) and (b)
in Fig.\ref{figure:Fig1}]. We list our numerical results on the
branching ratio for $B_c\rightarrow J/\psi K$ decays in Table
\ref{tab1}, where $\omega$ is the parameter in the wave function
of the $J/\psi$ meson(see Ref.\cite{SDY} for details). From the
numbers in Table \ref{tab1}, one can find that the branching ratio
of $B_c\rightarrow J/\psi K$ decays is sensitive to the parameter
$\omega$. The branching ratio of $B_c\rightarrow J/\psi K$ decays
is more sensitive to the quark masses, especially the c-quark's
mass, as shown in Table \ref{tab2}. Therefore the $B_c\rightarrow
J/\psi K$ decays provide a good platform to understand the wave
function of the $J/\psi$ meson and the constituent quark model.
Besides the uncertainty from the parameter $\omega$ and the quark
masses, we find the uncertainty of the decay constant $f_{J/\psi}$
will bring about $7\%$ uncertainty to the branching ratio of
$B_c\rightarrow J/\psi K$ decays. We investigate the branching
ratio's dependence on the hard scale $t$ in Eq.
(\ref{eq:convolution1}), which characterize the size of
next-to-leading order contribution. The branching ratio is shown
in Table \ref{tab3} with those uncertainties. By changing the hard
scale $t$ from $0.8t$ to $1.2t$, we find the branching ratio for
$B_c\rightarrow J/\psi K$ decays changes little as shown in Table
\ref{tab3}, that mean the uncertainties in the next-to-leading
order contributions can be neglected for this decay mode. The
value of $\Lambda_{QCD}$ also affect the branching ratio of
$B_c\rightarrow J/\psi K$ decays, we have taken
$\Lambda_{QCD}=250MeV$ at $N_f=4$ as our default input. A recent
determination of $\Lambda_{QCD}$ gives
$\Lambda_{QCD}^{N_f=4}=234\pm26MeV$\cite{Blumlein}. Our result is
$Br(B_c\rightarrow J/\psi K)=1.70\times10^{-4}$, if
$\Lambda_{QCD}=208MeV$; and $Br(B_c\rightarrow J/\psi
K)=2.02\times10^{-4}$, if $\Lambda_{QCD}=260MeV$. Considering the
uncertainties from the input parameters, our results on the
branching ratio of $B_c\rightarrow J/\psi K$ decays are generally
in the range of $(1-3)\times 10^{-4}$. The large branching ratio
and the clear signals of the final states make the measurement of
$B_c\rightarrow J/\psi K $ easily at the LHC-b experiments.

\begin{table}[htb]
\begin{center}
\begin{tabular}[t]{r|c|c|c}
\hline \hline
      & $\omega=0.5GeV$ &  $\omega=0.6GeV$ & $\omega=0.7GeV$ \\
  \hline
  $Br(B_{c}^{+}\rightarrow J/\psi K^{+})$
  & $1.52$  & $1.96$  & $2.44$\\
\hline \hline
\end{tabular}
\end{center}
\caption{Branch ratios in the unit $10^{-4}$ for different
$\omega$} \label{tab1}
\end{table}

\begin{table}[htb]
\begin{center}
\begin{tabular}[t]{r|c|c|c}
\hline \hline
      & $M_c=1.72GeV$ &  $M_c=1.82GeV$ & $M_c=1.92GeV$ \\
  \hline
  $Br(B_{c}^{+}\rightarrow J/\psi K^{+})$
  & $3.80$  & $1.96$  & $1.19$\\
\hline \hline
\end{tabular}
\end{center}
\caption{Branch ratios in the unit $10^{-4}$ for different $M_c$
using $\omega=0.6GeV$ and $M_b=5.2GeV$} \label{tab2}
\end{table}

\begin{table}[htb]
\begin{center}
\begin{tabular}[t]{|r|c}
 \hline     \hline
 Scale    & $Br(B_{c}^{+}\rightarrow J/\psi K^{+})$   \\
\hline 0.8 $t$   & $2.10 \times 10^{-4}$  \\
\hline
 $t$      & $1.96\times 10^{-4}$     \\
\hline 1.2 $t$    & $1.91\times 10^{-4}$  \\
\hline  \hline
\end{tabular}
\end{center}
\caption{Branching ratio using $\omega=0.6GeV$ with uncertainties}
\label{tab3}
\end{table}

\begin{table}[htb]
\begin{center}
\begin{tabular}[t]{r|c|c|c|c|c|c|}
\hline \hline
      & \cite{IKS} &  \cite{KKL,Kiselev} & \cite{CC} & \cite{Anisimov,Colangelo} &
\cite{EFG} &
      \cite{HMV} \\
  \hline
  $Br(B_c\rightarrow J/\psi K)$
  & $1.3\times 10^{-4}$  & $1.1\times 10^{-4}$  & $1.4\times 10^{-4}$ & $0.7\times 10^{-4}$ & $0.5\times 10^{-4}$ & $0.8\times 10^{-4}$   \\
\hline \hline
\end{tabular}
\end{center}
\caption{Branching ratio for $B_c\rightarrow J/\psi K$ decay in
the literature.} \label{tab4}
\end{table}

In the literature, there already exist a lot of studies on
$B_{c}\rightarrow J/\psi K$
decays\cite{IKS,KKL,Kiselev,CC,Anisimov,Colangelo,EFG,HMV}, we
show their results in Table \ref{tab4}. All the previous works on
this decay model were based on naive factorization, which, as
expected, can be quite accurate for the $B_c$ meson, since the
quark-gluon sea is suppressed in the heavy
quarkonium\cite{Kiselev}. In the naive factorization approach, the
decay amplitude can be expressed in terms of the hadronic
transition form factors and of the leptonic decay constants. For
the $B_c\rightarrow J/\psi$ transition form factors, there is a
large difference among the previous works\cite{SDY}. The authors
of Ref.\cite{KKL,Kiselev} calculated the form factors in the frame
work of QCD sum rules, as argued by the authors, the sum rule
estimates of form factors are taken at zero transfer squared,
while the dependence on $q^{2}$ is beyond the reliable accuracy of
the method.  The authors of Ref.\cite{Anisimov,Colangelo,EFG,HMV}
presented their calculations based on the constituent quark model,
where, the $B_c$ decay form factors can be expressed through the
overlap integrals of the meson wave functions. In Ref.\cite{CC},
Chang and Chen used a very complicated approach to evaluate the
weak current matrix elements. It is expected that the approach is
available as long as the mesons in the initial and final states
are of weak binding, for example, for the decay $B_{c}\rightarrow
J/\psi K$. After updating the model parameters, Ivanov $et$ $al$.
studied exclusive nonleptonic and semileptonic decays of the $B_c$
meson within a relativistic constituent quark model in
Ref.\cite{IKS}. For the branching ratio of $B_c\rightarrow J/\psi
K$ decays, their results are close to our predictions. In general,
the results of the various model calculations on $B_{c}\rightarrow
J/\psi$ decays are of the same order of magnitude. Eventually, the
variations among the theoretical predictions should be discerned
by experimental results of the $B_c$ decays, where, one can learn
much about the decay mechanism.

 This work is supported by the National Natural Science Foundation
of China under Grant Nos. 10847157 and 10575083. We would to thank
Dr. M.-Z. Zhou and Dr. J.-F. Sun for valuable discussions.

\end{document}